\documentclass[preprint,pra,letterpaper,showpacs,superscriptaddress,floatfix]{revtex4-1}
\usepackage{graphicx,psfrag,bbm,latexsym,color,dcolumn,bm,dsfont,bbm,color,mathrsfs,bbold,latexsym,amsmath,amsfonts,amssymb}

\newcommand{\beq}{\begin{equation}}
\newcommand{\eeq}{\end{equation}}

\newcommand{\beqa}{\begin{eqnarray}}
\newcommand{\eeqa}{\end{eqnarray}}

\begin{document}

\title{Gr\"obner Bases for Finite-temperature Quantum Computing and their Complexity}
\author{P.R. Crompton}
\affiliation{Department of Applied Mathematics, School of Mathematics, University of Leeds,  Leeds, LS2 9JT, U.K.}
\vspace{0.2in}
\date{\today}

\begin{abstract}
{Following the recent approach of using order domains to construct Gr\"obner bases from general projective varieties, we examine the parity and time-reversal arguments relating de Witt and Lyman's assertion that all path weights associated with homotopy in dimensions $d\leq 2$ form a faithful representation of the fundamental group of a quantum system. We then show how the most general polynomial ring obtained for a fermionic quantum system does not, in fact, admit a faithful representation, and so give a general prescription for calcluating  Gr\"obner bases for finite temperature many-body quantum system and show that their complexity class is BQP. }
\end{abstract}


\maketitle

\section{Introduction}
One of the most important open problems in quantum theory is determining the connection between Artin's representation of the Braid group \cite{-2} and the representation of physical many-body quantum systems. The anyon conjecture - that the representation theory of the closed paths of elementary identical particles (anyons) is more important for understanding the excitations of many-body quantum systems than the representation theory of the elementary particles themselves \cite{-1}\cite{-3}, has been used to promote topological proposals for quantum computation \cite{12}\cite{13}\cite{2}. However, more fundamentally, because the path weights associated with these anyons form a one-dimensional unitary representation of the fundamental group of the quantum system \cite{-4}\cite{0} (which in two physical dimensions forms a representation of the Braid group \cite{-5}), this frames the question as to whether in the broadest sense quantum systems are integrable, and have exact solutions. Specifically, the question we seek to address in this article is whether the above foundational approaches for quantum theory are restricted to systems which have an underlying Clifford algebra \cite{-5}, as was the case with original cases considered for $SO(3)$ and $SO(2)$ \cite{-3}\cite{0}, or indeed are restricted to systems in which the Clifford group forms a stabilizer \cite{2} or normalizer \cite{1} of the fundamental group of the quantum system or, if in fact, there is a much more general prescription for constructing algebras for topological quantum computing, via Gr\"obner bases \cite{27}, which is more widely applicable for finite temperature systems. 


\section{Clifford Algebra}

To cast our question into an explicit form, the configuration space of $N$ identical particles in a $D–$ dimensional Euclidean space can be written as \cite{-4}, 

\begin{equation}
M^{D}_{N} = \frac{(\mathbb{R}^{D})^{N}-\Delta}{S_{N}}
\end{equation}

where $S_{N}$ is the permutation group, and $\Delta$ are the fixed points generated by the action of $S_{N}$ on $(\mathbb{R}^{D})^{N}$. The propagator for a configuration of the physical many-body quantum system $a \in M^{D}_{N}$ at time $t_{a}$ to evolve into a configuration $b$ at $t_{b}$ can then defined via the path integral
\begin{equation}
K(b, t_{b}; a, t_{a}) = \int^{q(t_{b})=b}_{q(t_{a})=a} Dq \, e^{i\mathcal{S}}
\end{equation}

where $q(t)$ is the path traced out by an elementary particle in space, and $\mathcal{S}$ is the action \cite{0}. By re-ordereding this expression through the homotopy class of closed paths between $a$ and $b$, denoted $\pi(M^{D}_{N} , a, b)$, it can be noticed \cite{0} that the weights form a one-dimensional unitary representation of the fundamental group associated with the multiplication of closed paths,
\begin{equation}
K = \sum_{\alpha \in \pi} \chi(\alpha)K^{\alpha} \quad, \quad \chi(\alpha)\chi(\beta) = \chi(\alpha.\beta)
\end{equation}

Therefore, the difficulty in determining the connection between Artin's representation of the Braid group \cite{-2} and the representation of  physical many-body quantum systems essentially lies in connecting the deformation of the homotopy classes which categorise the closed anyon paths \cite{-3} with the topology of physical quantum systems. For example, if we vary the temperature of a many-body quantum system or interact with the quantum system in (3) to make a measurement in some non-adiabatic way, such that $b\rightarrow b'$, this can change the homotopy classes of the system in a nontrivial way \cite{24}, destroying the representation in (3). 

Our initial focus will be Fauser's work \cite{4}\cite{25} on identifying the grading assumptions associated with obtaining Clifford algebras from generalisations of the Chevalley deformation of a general polynomial ring, and we will show what normal ordering assumptions arise from this, and how these lead to the construction of a natural Gr\"obner basis for computing proposals for finite temperature quantum systems. The path integral representation of the propagators in (2) and their associated algebras (specifically, the generating functionals used to describe the Schwinger-Dyson hierachy of quantum field theories) are constructed by defining the (fermionic) Schwinger sources $j_{I}$ and their duals $\partial_{I}$ \cite{4} via
\begin{equation}
\{\partial_{I_{1}}, \partial_{I_{2}}\} = 0, \quad \{j_{I_{1}}, j_{I_{2}} \}= 0, \quad \{\partial_{I_{1}}, j_{ I_{2}}\} = \delta_{I_{1}I_{2}}
\end{equation}
where $I$ are the set of all relevant quantum numbers. As Fauser shows, a functional Fock space can then be used to define these generating functionals, which are of the form 
\begin{equation}
\left| \mathcal{T} (j ) \right\rangle := \sum_{n=0}^{\infty} \frac{i^{n}}{n!} \tau_{n}(1,...,n) j_{1}...j_{n}\left|0\right\rangle_{F}, \quad \partial_{I}\left|0\right\rangle_{F}=0 
\end{equation}
where $\tau_{n}$ are the time-ordered correlation functions. The exterior algebra of the linear space spanned by these sources, $V = \langle j_{I} \rangle$, is then formally constructed as the polynomial ring in these anti-commuting Schwinger sources \cite{4}
\begin{equation}
\{j_{I_{1}}, j_{I_{2}}\}_{+} = 0, \quad \bigwedge V = \mathbb{C} \oplus V \oplus V \wedge V \oplus ... , \left|\mathcal{T} (j ) \right\rangle \in V
\end{equation}

It is important to note that although the exterior algebra of this general quantum system has now been defined with respect to some number of identical fermionic particles, anyons are not yet necessarily defined for this general quantum system because the equivalence classes for the path weights in (2) have yet to be constructed. Moreover, the exterior algebra in (6) will be generally defined through the grade projection operators $\langle  . . . \rangle_{r} : \wedge V \rightarrow \wedge^{r} V$. Hence, if we now attach an explicit (multi-linear) form and grading to this algebra this step will be the most physically important for defining the general finite temperature quantum system, since it will define the time-reversal symmetry of the system. An explicit time-reversal symmetry operator $T$ for the system can be defined through the general left and right operations applied to (5) \cite{25} via
\begin{equation}
\psi^{+} = \partial_{I} + \frac{1}{2}T_{IL}j_{L}\wedge, \quad \psi^{-} = \partial_{I} - \frac{1}{2}T_{LI}j_{L}\wedge\,,
\end{equation}
where the time-ordering of a specific state $\psi$ is defined through the transition element
\begin{equation}
\left| \mathcal{T} (j,r) \right\rangle := \sum_{n=0}^{\infty} \frac{i^{n}}{n!} \tau_{n}(1,...,n|r) j_{1}\wedge...\wedge j_{n}\left|0\right\rangle_{F},\quad
\tau_{n}(1,...,n|r) = \left\langle 0 | \mathcal{T} (\psi_{I_{1}}, ... \psi_{I_{n}}) | r \right\rangle
\end{equation}
The anyon proposal \cite{-1}\cite{-3} can be read, therefore, in this context as a prescription for defining time-reversal symmetry where the above transition matrix elements are homomorphic to the path weights $\tau_{n}(1,...,n|r) \rightarrow \chi(r)$ \cite{0}. Although, it is again important to note that it is only if the representation of the path weights in (3) is trivial that this symmetry remains unbroken. 

In order to then decide which multi-linear forms and gradings are important for describing physical many-body quantum systems we can restrict our attention (initially) to quantum systems in which we only have antisymmetric tensors defined for (6). This reduces the exterior algebra in (6) to the tensor algebra
\begin{equation}
T(V)=\mathbb{C} \oplus V \oplus (V \otimes V) \oplus ... = \oplus_{r}T^{r}(V) = \oplus_{r}\otimes^{r} V
\end{equation}
Forming the quotient of this tensor algebra with one of the following two ideals will then define the well-known Grassmann or Clifford algebras
\begin{eqnarray}
\mathcal{I}_{\wedge} & = & \{ a\otimes x \otimes x \otimes b \,\,| \,\,a,b \in T(V), x \in V\} \\ \mathcal{I}_{Cl} & = & \{ a\otimes (x\otimes y + y\otimes x) \otimes b) -2\eta(x,y)a\otimes b\,\,|\,\,a,b \in T(V), \,\,x,y,\in V\}
\end{eqnarray}
However, clearly this approach is not very instructive for (10) since we are reducing the polynomial ring in (6) to a factor algebra where the grade projection is onto a trivial center, but for more subtle reasons, we now argue, neither is (11). Formally, the Clifford algebra $Cl$ associates a quadratic form $\eta$ to the associative algebra generated by the elements $\{\Gamma_{l}\}^{D}_{l=0}$ which satisfies  $\{\Gamma_{i}, \Gamma_{j} \} = 2\eta_{ij}\mathbf{1}\quad i,j = 1, . . . ,D$, and it is a way to represent the square root of the signature $-\eta(x)\mathbf{1}$ where $\mathbf{1}$ is the unit of the algebra $Cl$. This minus sign is therefore the same one which appears in the generalised (time) symmetry operation in (7). However, the subtle point is that we also want the quantum states in (8) to have a similar ordering to (7) in the systems dual algebra (bialgebra) $[\,\bigwedge V]^{*}$, because we are anticipating the path weights in (3) form a one-dimensional unitary representation of the fundamental group of the system \cite{0} 
and these are homomorphic to the transition matrix elements. Hence, the naturally $\mathbb{Z}_{2}$-graded Clifford algebra can only represent a fermionic quantum system of this form which is either planar in time ($b\rightarrow b'$ invariant) or in space ($a\rightarrow b$ invariant), otherwise the parity symmetry operator $P$ defined on the dual space by
\begin{equation}
\psi_{+} = j_{I} + \frac{1}{2}P_{IL}\partial_{L}\wedge, \quad \psi_{-} = j_{I} - \frac{1}{2}P_{LI}\partial_{L}\wedge\,
\end{equation}
\begin{equation}
\left| \mathcal{P} (I,r) \right\rangle := \sum_{n=0}^{\infty} \frac{i^{n}}{n!} \rho_{n}(1,...,n|r) \partial_{I_{1}}\wedge...\wedge \partial_{I_{n}}\left|0\right\rangle_{F},\quad
\rho_{n}(1,...,n|r) = \left\langle 0 | \mathcal{P} (\psi_{j_{1}}, ... \psi_{j_{n}}) | r \right\rangle
\end{equation}
will generate a different grading for the time-ordered projection of the quantum system defined in (8), or vice-versa \cite{23}. Our basic observation, which extends Fauser's analysis, is to note that  a general exterior algebra which is used to represent a quantum system (and which incorporates the Clifford algebra) must either be parity or time symmetry symmetric (with these symmetries defined via (8) and (12)), or the representation in (3) will be unfaithful. 

\section{normalizers and stabilizers}

Recent approaches to defining a representation theory for physical many-body quantum systems have focused on treating the Clifford algebra as a stabilizer \cite{2} or normalizer \cite{1} within some larger algebra, to address the problems identified above. The general procedure of Cliffordization, for example, can be understood from Fauser as the generalized Chevalley deformation in (17) and (12) that is composed from the left or right contraction of the exterior product of Grassmann-Hopf algebras \cite{2555}, although crucially the antipode is well-defined for such a system. Our new approach, which follows similar lines, is to construct the polynomial ring in (6) directly from a quotient ring defined by an ideal quotient of $\mathcal{I}_{Cl}$. Where, formally, for $\mathcal{I}$ and $\mathcal{J}$ two ideals of a commutative ring $R$, the ideal quotient $(\mathcal{I} : \mathcal{J})$ is defined as the set $\mathcal{I}:\mathcal{J} = \{ r \in R | r\mathcal{J} \subset \mathcal{I} \}$, which we will show can be extended to form a natural Gr\"obner basis for computing the polynomial ring and, moreover, allows us to treat more general grade projections. However, before formally defining the deformations that define this quotient ideal it is important to formalise the role of anti-particle representations of the quantum field theory in (2). It is possible, for example, to define the Clifford group as the normalizer of any Pauli operator $\mathcal{P}_{n}$ in $U(2n)$, i.e. via a unitary embedding \cite{1} 
\begin{equation}
Cl_{n} = \left\{ U \in U(2n) |\,\,U\mathcal{P}_{n}U^{\dagger} = \mathcal{P}_{n}\right\}
\end{equation}
However, there is a problem in defining the quantum system in (2) from this construction which is introduced by the $\mathbb{Z}_{2}$-grading of the Clifford algebra, namely, the resultant quantum system must parity invariant for the representation in (14) to be faithful, as we have just identified. Similarly, this approach is applied more widely in quantum computing \cite{12}\cite{13} for defining the quantum system given by the unitary embedding of the Fock space defined in (5) within a qubit space $\mathcal{B}$, defined as $\mathbb{C}^{2}$ endowed with the standard basis $\left\{\left|0 \right\rangle, \left|1 \right\rangle \right\}$. Specifically in \cite{2}, this unitary embedding can be defined as $J : \mathcal{H} \rightarrow \mathcal{B}^{ \otimes n}$, where $\mathcal{H} = \mathcal{H}_{0} \otimes \mathcal{H}_{1}$ is the Fock space defined in (5) split into the subspaces corresponding to an even and odd number of identical particles, where the relation between the representation of two operators in an algebra $L$ is given by
\begin{equation}
U \in L(\mathcal{H}), \quad U' \in L( \mathcal{B}^{\otimes n} ) \quad\quad \quad JU=U'J
\end{equation}
Whilst (15) now admits the possibility of defining anyons via the stabilizer subgroups of $J$, because the representation of $J$ must be faithful in (15), the quantum system it defines must again by parity invariant, following the same argument. 

The important question for finding a general representation theory for a finite-temperature quantum system is, therefore, whether the existence of a parity symmetry is a necessary and sufficient condition for the anti-particle representation to exist. One general means to ensure the anti-particle representation can be explicitly constructed from the particle representation, for example, is to construct a Clifford-Hopf algebra \cite{21} which is simultaneously an algebra and coalgebra equipped with a counit and comultiplication and which, crucially, also has an antipodal map. The Clifford-Hopf algebra $ClH$ is the associative algebra defined by the generators $\Gamma_{l} (l = 1, . . . ,D+1)$ and central elements $E_{l} (l =1, . . . ,D)$ via
\begin{eqnarray}
& & \Gamma_{i}^{2}=E_{i}, \quad
\Gamma_{D+1}^{2}=\mathbf{1}, \quad
\{\Gamma_{i}, \Gamma_{j} \} = 0 \quad i \neq j\\
& & [E_{i},\Gamma_{j}] = [E_{i},\Gamma_{D+1}] = [E_{i},E_{j}] = 0 \quad \forall\,\, i,j
\end{eqnarray}
in addition to the following relations for the comultiplication map $\Delta$, counit map $\epsilon$ and antipodal map $S$ which ensure that these three maps are algebra morphisms

\begin{equation}
\begin{array}{lll}
\Delta(E_{i}) = E_{i} \otimes\mathbf{1} + \mathbf{1} \otimes E_{i}, & 
S(E_{i}) = -E_{i}, & 
\epsilon(E_{i}) = 0 \\
\Delta(\Gamma_{i}) = \Gamma_{i}\otimes\mathbf{1} + \Gamma_{D+1}\otimes\Gamma_{i},  & S(\Gamma_{i}) = \Gamma_{i}\Gamma_{D+1},  & \epsilon(\Gamma_{i}) = 0 \\
\Delta(\Gamma_{D+1}) = \Gamma_{D+1}\otimes\Gamma_{D+1}, & S(\Gamma_{D+1}) = \Gamma_{D+1},& 
\epsilon(\Gamma_{D+1}) = 1
\end{array}
\end{equation}

The important relation for parity symmetry is, therefore, the one in the middle column which tells us that the sign of the central elements is flipped by the antipodal map. Although we can now define the anyons of (3) explicitly via the central elements of this algebra, the Clifford-Hopf algebra is still explictly parity invariant. This can be seen directly from (16) and (17) by noticing \cite{3} that for even $D$ the central elements are Casimirs of $ClH$, as is the product $\prod_{i=1}^{D+1} \Gamma_{i}$. Hence, for even $D$ the central elements are given by $E_{i}= \eta_{ii}$ and the irreducible representations of $ClH$ are therefore isomorphic to those of the Clifford algebra $Cl$. Similarly, for odd $D$ the irreducible representations of $ClH$ are related to those of $Cl$, but defined in one dimension higher. We are therefore led to the conclusion that in order to treat the general algebra of the polynomial ring in (5), applicable to a quantum system where $b\rightarrow b'$ in (2),  we must weaken the construction of the above antipode such that it is graded, via some form of quantum group deformation: general finite-temperature quantum systems cannot be solved exactly by stabilizer or normalizer constructions of the Clifford algebra.

\section{Quantum Deformation of a Toric Variety}

\subsection{Toric Variety}
Our approach will now be to identify the variety that is associated with the ideal quotient $(\mathcal{I} : \mathcal{J})$ that we are seeking. Our first step will be to picture the parity symmetry operator of (12) as being equivalent to introducing light cone coordinates, with the Lorentzian signature $(D,1)$, for the Clifford ideal $\mathcal{I}_{Cl}$ in (11) 
\begin{equation}
 ds^{2} = - dt^{2} + \delta_{ij}dx^{i}dx^{j}  \quad \rightarrow \quad ds^{2} = -2dx^{ +} dx^{-} + \delta_{ij}dx_{i}dx_{j}
\end{equation}
where $i,j = 1,...,D-1, x^{+}=\frac{1}{\sqrt{2}} (t+x)$  and $x^{-}=\frac{1}{\sqrt{2}}(t-x)$, and to identify the variety associated with this coordinate system. Formally, the variety that is associated with this coordinate system is defined for a general polynomial ring (defined in the $D + 1$ variables $x_{i}$) via the homogeneous coordinate ring $\mathbb{X}$ \cite{16}\cite{300}.  This can be written as the quotient ring
\begin{equation}
\mathbb{X} = \mathbb{C} [x_{0}, x_{1}, x_{2}, ..., x_{D}]/\mathcal{I}
\end{equation}
where $\mathcal{I}$ is a homogeneous ideal, and it is important to note that by specifying that the above ideal is homogenous an implicit form of grading is assumed, which allows us to focus (initially) on the parity invariant case. To define this ideal and grading further, we let $\Lambda$ be a lattice of rank $D$ and $\check{\Lambda}$ its dual lattice, where $\Lambda_{\mathbb{R}} = \Lambda \otimes_{\mathbb{Z}} \mathbb{R}$ is called a regular $k$-dimensional cone defined as a convex subset $\sigma \subset \Lambda_{\mathbb{R}}$ \cite{16} via
\begin{equation}
\sigma = \mathbb{R}_{\geq 0}\langle x_{i} \rangle^{k}_{i=1} = \left\{ \sum_{i=1}^{k} a_{i}x_{i} \,\,| \,\,a_{i} \in \mathbb{R}_{\geq 0} \right\}
\end{equation}
where  $\{x_{1}, . . . , x_{k}\}$ is some subset of $\Lambda$ which can be extended into a basis. The set of all cones can then be combined to form the following set, known as a complete regular fan 
\begin{equation}
s(D) = \{\sigma_{I}\}_{I \subset \{1,...,D+1\}}, \quad \sigma_{I} = \mathbb{R}_{\geq 0} \langle x_{j} \rangle_{j \in I}
\end{equation}
Finally, to form the projective space associated with the $D$-dimensional light cone coordinates for (16) we can identify the affine varieties associated with each $\sigma$ and can then glue these together \cite{300}, since these varieties  $\mathbb{A}_{\sigma} = {\rm{spec}} \,\, \mathbb{C}[\check{\sigma}]$  form the natural inclusion  $\mathbb{A}_{\sigma} \hookrightarrow \mathbb{A}_{\sigma' < \sigma}$, where  
$\check{\sigma} = \{ \check{\lambda} \in \check{\Lambda} \,\, | \,\, \langle \check{\lambda}, v\rangle \geq 0, \,\, \forall v \in \sigma \}$ defines the dual cone. It follows that the variety describing the light cone coordinates in (16), which is known as a toric variety, is defined via 
\begin{equation}
\mathbb{T}_{S} = {\rm{spec}} \,\,\mathbb{C} [ \check{\Lambda} ] 
\simeq {\rm{spec}} \,\,\mathbb{C} [  x_{i}^{\pm}]^{D}_{i=1} = (\mathbb{C}^{\times})^{D}
\end{equation}
where $\mathbb{C} [  x_{i}^{\pm}]^{D}_{i=1}$ is the homogeneous coordinate ring for the algebraic torus $(\mathbb{C}^{\times})^{D}$. Therefore, although we have made progress towards defining a suitable ideal quotient we still need to introduce further grading in order to construct the polynomial ring in (6), since it follows from (20) that the polynomial ring in (6) is only the homogeneous coordinate ring of the above projective space not of the space associated with the exterior algebra. 

\subsection{Quantum Deformation}

The quantum deformation of the above toric variety in (23) can be defined, however, via the quantization of the homogeneous coordinate ring of the above algebraic torus. Defining $q = (q_{ij} ) \in M_{D}(\mathbb{C}^{\times})$ as the multiplicatively antisymmetric matrix, such that $q_{ii} = 1$ and $q_{ji} = {q_{ij}}^{-1}\,\, \forall \,\, i, j$, the quantum torus over $\mathbb{C}$ is defined as the algebra \cite{15}
\begin{equation}
\mathcal{O}_{q}((\mathbb{C}^{\times})^{D}) := \mathbb{C} \langle x^{\pm 1}_{1}, ... , x^{\pm 1}_{n} \,\,| \,\,x_{i}x_{j} = q_{ij}x_{j}x_{i} \,\, {\rm{for}} \,\, {\rm{all}} \,\, i,j \rangle
\end{equation} 
The important point about this construction, which makes it the natural language for talking about the time and parity symmetries of the quantum states in (2), is that unlike (20) the natural geometric objects associated with the polynomial ring in (6) are not its set of maximal ideals, but rather its set of prime ideals. To see this explicitly, from (21), if $x \in \mathbb{A_{\sigma}}$ (the affine variety) corresponds to a maximal ideal $\mathcal{I}_{max}$, then $x \mapsto  \mathcal{I}_{max} \subset \mathbb{A_{\sigma}}$ defines a one-to-one correspondence between the $x$ and the maximal ideals of $\mathbb{A_{\sigma}}$.
However, if we now look at (21) and construct the ring homomorphism $S: a \rightarrow b$ as a regular map between the ideal $x \subset b$ and some ideal $x \subset a$ the inverse image of the maximal ideal does not have to be maximal, but can belong to some subvariety $\mathbb{A}_{\sigma'}$, and is hence prime. This is essentially the geometric equivalent of the difference between a reducible and irreducible representation.

\subsection{$\mathbb{Z}_{2}$-grading and the Braid group}

Our approach so far has been to identify and match the properties of the quotient algebras of the generic exterior algebra in (6) with the desirable time and parity symmetry properties of physical many-body quantum systems, specifically those at finite-temperatures. In particular, we have focused on identifying why Clifford algebras and their coalgebras are important for the representation of quantum many-body systems \cite{0}\cite{-5}, but also why stabilizer and normalizer formulations are necessarily limited \cite{2}\cite{1} in regards the whether the existence of a parity symmetry is a necessary and sufficient condition for the anti-particle representation of quantum system to exist. Having now identified the broadest possible suitable deformation of a variety to form the coordinate system for this quotient algebra in (24) we will now reverse this approach to see how coordinate rings for (24) relate to the one dimensional unitary representation of the Braid group in (3). So, starting with some algebra $X$ defined on  $\mathbb{C}$, and some abelian group $G$, the 2-cocycle defined by $\chi : G \times G \rightarrow \mathbb{C}^{\times}$ gives the associated $G$-graded vector space a new associative multiplication rule \cite{15}. The new $G$-graded algebra that is generated by this multiplication is known as the twist of $X$, defined as $Y$, and its associative multiplication rule is explicitly of the form 
\begin{equation}
x_{i} * x_{j} = \chi(\alpha, \beta) \, x_{i} x_{j} 
\end{equation}

where $x_{i} \in X_{\alpha}$ and $x_{j} \in X_{\beta}$. Note, we have used the same notation for the 2-cocyle as the path weights defined for fixed homotopy in (3) to make the connection between them explicit. 
Relating this construction back to the Clifford-Hopf algebra definition in (15) if we now assume that $\chi$ is an alternating bicharacter (which is the same choice, rather than restriction, that we used for (24)) then the maximal and prime spectra, respectively, of $X$ and $Y$ are given by
\begin{eqnarray}
{\rm{max}}_{\alpha} X&=&{\rm{max}}X \cap {\rm{spec}}_{\alpha}X, \quad{\rm{spec}}_{\alpha} X = \{ P \in {\rm{spec}} \, X \,\,| \,\,x_{i} \in P \Leftrightarrow i \in \alpha \} \\
{\rm{max}}_{\alpha} Y&=&{\rm{max}}Y \cap {\rm{spec}}_{\alpha}Y,\,\,\, \quad {\rm{spec}}_{\alpha} Y = \{ P \in {\rm{spec}} \, Y\,\, |\,\, y_{i} \in P \Leftrightarrow i \in \alpha \}
\end{eqnarray}
where $x_{1}, . . . , x_{n}$ are the homogeneous generators of $X$, and $y_{i} \, \in \, Y$ for $i = 1, ... \, n$. Although the definitions in (26) and (27) look fairly obvious and innocuous, the implications are subtle. Namely, that a general $G$-graded space of this form cannot posses an antipode because, as we have seen in (24), it is the prime and not the maximal spectra of $Y$ which is important: the inverse image of the maximal spectra does not have to be maximal. This is a known result for the Clifford-Hopf algebra \cite{21}, for example, which can be seen explicitly via the biconvolution defined by $Cl(\eta, \xi)$, where the algebra is defined via $Cl(V, \eta \in V^{*} \otimes V^{*})$ and the coalgebra is defined via $Cl(V, \xi \in V\otimes V)$. In this case, it is only for a special choice of biconvolution, namely when the biconvolution is antipodal, that $Cl(\eta, \xi)$ is isomorphic to the Clifford-Hopf algebra. 

The key point for us is that creating a homogeneous coordinate ring of the form of (23) for the polynomial ring in (6) does not mean that the twist in (25) is mapped to the same set of homogenous coordinates. Therefore,
the general result for a finite temperature quantum system is that although the path weights in (3) form a one-dimensional unitary representation of the fundamental group of the system (defined via $x_{i} \in X_{\alpha}$ and $x_{j} \in X_{\beta}$), this representation is not generally faithful, because 
\begin{equation}
x_{i} * x_{j} = \chi(\alpha, \beta) \, q_{ij}x_{j}x_{i} \neq  \chi(\beta, \alpha) \, q_{ji}x_{i}x_{j} = x_{j} * x_{i} 
\end{equation}
ie. the general quantum deformed coordinates in (24) do not form an associative group multiplication rule in conjunction with the 2-cocyle. Nonetheless, even though the representation is unfaithful, the above 2-cocyle does form a one-dimensional unitary representation of the Braid group $B_{n}$ if the $x_{i}$ satisfy the following relations \cite{13}
\begin{equation}
x_{i} * x_{i+1} * x_{i} = x_{i+1} * x_{i} * x_{i+1} , \quad 
x_{i} * x_{j} = x_{j} * x_{i}  \quad |i-j| >1
\end{equation}
Central to this definition is notion of the general closure of the braid $x$ which is defined for the two braids $x_{i}, x_{j} \in B_{n}$ and the nonnegative integers $\alpha$ and $\beta$, satisfying $2\alpha + \beta = n$, by
$\chi(\alpha,\beta)^{(x_{i},x_{j})}(x)$. Again, we have used the same notation for the braid closure and 2-cocyle to make their connection explicit. A simple way to satisfy (28) and the second relation in (29) is to make the quantum deformation matrix $q_{ij}$ an involution, for example, by chosing $q_{ij}$ to be a Hadamard matrix. If $q$ is chosen in this way then the $|i-j|>1$ condition is automatically satisfied from our previous choice of $q$ as an alternating bicharacter ($q_{ii}=1$). Similarly, we can satisfy the first relation in (29) by using just the associativity properties of the alternating bicharacter of the quantum deformation matrix and 2-cocyles
\begin{eqnarray}
(x_{i} * x_{j}) * x_{k} & = & \chi(\alpha, \beta) \chi(\beta,\alpha) \, x_{i} x_{j} x_{k} = q_{jk} q_{ik} \,\, x_{k}x_{i}x_{j}  = q_{ji, \,k} \, x_{k}x_{i}x_{j}
\end{eqnarray}

The most limiting assumption for defining a general quantum system, therefore, is to choose $q$ as an involution. Morevover, it is only necessary to make this very specific choice for the quantum deformation if we are dealing with the prime spectra rather than the maximal spectra, where we do not have the inclusion $X_{\alpha} \hookrightarrow  X_{\beta}$. Hence, for a finite temperature quantum system if we do want to probe this general inclusion, via $\chi(\alpha,\beta)^{(x_{i},x_{j})}(x)$, we need to identify the ideal quotient of the homogeneous generators in (26) and (27) in order to define the more general resolution of (28), rather than the specific (faithful) one where $q$ is an involution. Given the two ideals $\mathcal{I}$ and $\mathcal{J}=(x_{1},x_{2})$ this is defined via
\begin{equation}
\mathcal{I}: \mathcal{J} = (\mathcal{I}:(x_{1}))\cap (\mathcal{I}:(x_{2})) = \left(x^{-1}_{1}(\mathcal{I}\cap(x_{1}))\right) \cap \left(x^{-1}_{2}(\mathcal{I}\cap(x_{2}))\right)
\end{equation}
Hence, the intersection between the ideal $\mathcal{I}$ and $(x_{1})$ is given by
\begin{equation}
\mathcal{I}\cap (x_{1}) = t\mathcal{I} + (1-t)(x_{1})\cap \mathbb{C}[x_{1}, ..., x_{n}] 
\end{equation}
where $t\in\mathbb{C}$ \cite{27}. Therefore, although we cannot generally find the construction of the anti-particle representation of this most general many-body quantum system at finite temperature, without considerable effort, we can easily identify that the basis functions which have no $t$ in them will generate $\mathcal{I}\cap (x_{1})$, and in fact these elements form an exact Gr\"obner basis, as will now show. The key point is that although usually we want the representation of the braid group to be formed via a known algebra $X$ (such as the Clifford algebra) and its twist $Y$ (for a faithful representation of the fundamental group of the quantum system \cite{0}), in (25), what we can see from the above is that it is more natural to braid the intersection between the primitive and maximal spectra of $X$, and $X$, as defined via the quotient ideal $(\mathcal{I}: \mathcal{J})$ in (31), which in this case we would define via the specific ideal in (11). This generalises the notion of the anyon in (3) to an object which can be used to form an exact representation of the Braid group of a general many-body quantum system at finite-temperatures, even when the unitary representation of the fundamental group of the system defined in (3) is unfaithful. 

\section{Gr\"obner Bases}


The simple properties of ideal quotients, which closely follow of those of complete regular fans we have discussed previously, are that for $\mathcal{I}$ and $\{ \mathcal{I}_{k} \}_{1 \leq k \leq r}$ being ideals in $\mathbb{C}[x_{1}, ... ,x_{n}]$, and $f$ a polynomial in $\mathbb{C}[x_{1}, ... x_{n}]$, it follows that 
\begin{equation}
\left( \bigcap_{k=1}^{r} \mathcal{I}_{k} \right) : \mathcal{I} = 
 \bigcap_{k=1}^{r} \left( \mathcal{I}_{k} : \mathcal{I} \right), \quad 
\mathcal{I} : \left( \sum_{k=1}^{r} \mathcal{I}_{k} \right) = 
 \bigcap_{k=1}^{r} \left( \mathcal{I} : \mathcal{I}_{k} \right)
\end{equation}
Hence the simple prescription given in \cite{27} for constructing a basis for an ideal quotient is; calculate $\langle g_{1}, ... g_{p} \rangle = \mathcal{I} \cap \langle f \rangle$, followed by $\langle g_{1}/f, ... g_{p}/f \rangle = \mathcal{I} : f $, then finally $\bigcap_{k=1}^{r} \left( \mathcal{I} : g_{p} \right)$. The further subtlety which makes this construction a Gr\"obner basis is the ordering of the monomials $x^{\alpha}$, specifically, identifying the largest monomial of $f$ with respect to the ordering which is denoted $LM(f,\prec)$. The definition of a Gr\"obner basis that follows \cite{27} is a set of polynomials $G$ where there exists some $g \in G$ such that $LM(g,\prec)$ divides $LM(f,\prec)$. We can then make the connection between the weight associated with each cone within a complete regular fan (the $a_{i}$ in (21)), and the support associated with each monomial order, following \cite{19}, by defining order domains which are a class of commutative ring. For this definition, the monomial ordering is chosen as follows; $x^{\alpha} \succ_{M,\tau} x^{\beta}$ if $M\alpha \succ M\beta$, or $x^{\alpha} \succ_{\tau} x^{\beta}$ if $M\alpha = M\beta$, where $M$ is an $r \times n$ matrix with entries in ${\mathbb{Z}}_{\geq 0}$ with linearly independent rows. It follows the order domain, defined as $\mathbb{F}_{q}[x_{1}, . . . ,x_{s}]/\mathcal{I}$, can be constructed from a general projective variety defined over a finite field (with poles at only one smooth $\mathbb{F}_{q}$ rational point), and moreover, that the Gr\"obner bases imply the existence of the toric deformations of this general variety \cite{26}. However, the important question for us, as separate from the construction of order domains, is if we can construct a general polynomial ring from these commutative rings, and from (33) the answer quite clearly is yes, because the quotient of an ideal quotient is also an ideal quotient in this scheme.

The general scheme for identifying the computational complexity of a quantum system presented in \cite{13} is to calculate a suitable trace invariant of a polynomial representation of the system mapped onto a commutative ring, which is then compared with a result derived using the additive approximation: if these concur within certain bounds, the quantum complexity class can be determined. This additive approximation is defined via the (polynomial) function $f: X \rightarrow \mathbb{R}$ and the normalization function $g: X \rightarrow \mathbb{R}$, and is defined such that the random variable $Z(x)$ is associated to $x \in X$ and $\delta >0$  is computable in polynomial time in the size of the problem instance $n$ and in $1/\delta$, where
\begin{equation}
{\rm{Pr}}\left( \left| \frac{f(x)}{g(x)} - Z(x) \right| \leq \delta \right) \geq \frac{3}{4}
\end{equation}
Moreover, if this result is combined with the Chernoff bound \cite{13}, it follows that the random variables $X,Y \in \{ \pm 1\}$ such that $\mathbb{E} [X+iY] = \langle x |U| x \rangle$ can be used to sample the complex unitary $U((\mathbb{C}^{2})^{\otimes n}) \equiv U_{n}$) in such a way that it is bounded from both below and above, which gives the following basic defintion of the complexity class BQP
\begin{equation}
|\langle x_{n}\,00...0 \,|\,U_{n}\,|\, 0...00 \,x_{n}\rangle|^{2} \left\{ \begin{array}{cc}
\leq 3/4 & x^{n} \in L \\
\geq 1/4 & x^{n} \notin L \end{array}\right.
\end{equation}
where $L\subseteq \{1,0\}^{*} = f^{-1}(1)$. Clearly, from the basic definition of monomial ordering, $LM(g',\prec) = 1 \,\, \forall \,\, g' \in G'$ can be computed from $G$ in polynomial time \cite{27}, but more importantly, degree reverse lexicographic ordering can be defined for the  construction of a general Gr\"obner basis via
\begin{equation}
-x_{1}^{\alpha_{1}} ... x_{n}^{\alpha_{n}} \prec_{DRL}  x_{1}^{\beta_{1}} ... x_{n}^{\beta_{n}}
\end{equation} 
with $\alpha = (\alpha_{1}, ..., \alpha_{n})$ and $\beta = (\beta_{1}, ... , \beta_{n} ) \in \mathbb{N}^{n}$, where  $\sum_{i=1}^{n} \alpha_{i} \geq \sum_{i=1}^{n} \beta_{n}$ and the right-most nonzero entry of $\alpha - \beta$ is negative. Hence, if $x^{d+1}_{n}$ divides the leading monomial of a polynomial it divides the entire polynomial and it follows that the corresponding Gr\"obner basis can, therefore, be computed in polynomial time in $d^{n}$ \cite{99}. Thus, even though, as we have argued, the anyon is a less useful object for a general many-body quantum system at finite-temperatures (because the unitary representation of the fundamental group of the system defined in (3) is unfaithful) computing a suitable projection onto a commutative ring for evaluating the bounds in (35) is implicitly polynomial in $d^{n}$. Moreover, because we have constructed an explicit polynomial ring, it also follows from (33) that the complexity class of computing the projection onto a commutative ring for (35) will be the same as the pure state case \cite{12}\cite{2}, since a suitable $\delta$ can be found via $(d/\delta)^{n} \leq n$.

To summarise, in this article we have identified, following Fauser \cite{4}, that  there are two sorts of grading on the general polynomial ring for the fermionic quantum system in (6), associated with time and parity reversal symmetries. Therefore, the unitary representation of the fundamental group of the system defined in (3) is unfaithful for this system \cite{0}, although we can construct construct a Gr\"obner bases \cite{27} for this polynomial ring via the toric and quantum deformations of a Clifford ideal, which have done defining an ideal quotient in (33). As we have shown this has allowed us to generalise the anyon concept \cite{-1}\cite{-3} to finite-temperatures, and form an exact representation of the Braid group, although the unitary representation of the fundamental group of the system defined in (3) is unfaithful \cite{0}. Moreover, we have shown, via the additive approximation, that the projections onto a commutative ring for evaluating the complexity bounds in (35) lead to this being BQP.

\end{document}